\documentclass{IEEEcsmag}

\usepackage[colorlinks,urlcolor=blue,linkcolor=blue,citecolor=blue]{hyperref}
\expandafter\def\expandafter\UrlBreaks\expandafter{\UrlBreaks\do\/\do\*\do\-\do\~\do\'\do\"\do\-}
\usepackage{upmath,color}

\usepackage{graphicx}
\usepackage{booktabs}
\let\labelindent\relax
\usepackage{enumitem}

\jvol{XX}
\jnum{XX}
\paper{8}
\jmonth{Month}
\jname{Publication Name}
\jtitle{Publication Title}
\pubyear{2024}

\renewcommand{\thesubsection}{C\arabic{subsection}.}
\newcommand{\customsubsection}[1]{%
  \refstepcounter{subsection}
  \subsection*{\thesubsection\space #1}
  \setlist[enumerate,1]{label*=\thesubsection\arabic*,before=\setcounter{enumi}{0}}
}

\begin{document}

\sptitle{Feature: Grey Literature Survey on the Release Readiness Criteria of Generative AI-based Software Products}

\title{A State-of-the-practice Release-readiness Checklist for Generative AI-based Software Products}

\author{Harsh Patel}
\affil{Queen's University, Canada}

\author{Dominique Boucher, PhD}
\affil{National Bank of Canada, Canada}

\author{Emad Fallahzadeh, PhD}
\affil{Queen's University, Canada}

\author{Ahmed E. Hassan, PhD}
\affil{Queen's University, Canada}

\author{Bram Adams, PhD}
\affil{Queen's University, Canada}

\markboth{FEATURE}{FEATURE}




\begin{abstract}\looseness-1
    This paper investigates the complexities of integrating Large Language Models (LLMs) into software products, with a focus on the challenges encountered for determining their readiness for release. Our systematic review of grey literature identifies common challenges in deploying LLMs, ranging from pre-training and fine-tuning to user experience considerations. The study introduces a comprehensive checklist designed to guide practitioners in evaluating key release readiness aspects such as performance, monitoring, and deployment strategies, aiming to enhance the reliability and effectiveness of LLM-based applications in real-world settings.
\end{abstract}

\maketitle

\chapteri{G}enerative AI, especially Large Language Models (LLMs), are increasingly being integrated into software products~\cite{hou2023large}, marking a significant shift in how companies approach product development and release. This shift is underscored by the potential economic impact of generative AI, which is estimated to contribute \$2.6 trillion to \$4.4 trillion annually to the global economy as per McKinsey's report\footnote{https://www.mckinsey.com/capabilities/mckinsey-digital/our-insights/the-economic-potential-of-generative-ai-the-next-productivity-frontier}.

Determining the release-readiness of a software product is complex, requiring a product's compliance with all user and safety requirements and successful passage of all quality assurance checks. For generative AI products, this readiness also involves evaluating model performance, ethical considerations, and the potential impact of outputs on users, compounded by the need to comply with evolving AI legislation across countries.

To some extent, some of the challenges of ensuring the production-readiness of \emph{generative} AI-based products in real-world settings are still the same as for systems leveraging \emph{traditional} machine learning (ML) systems. These challenges include non-determinism, due to the inherent variability in AI's decision-making processes, and the difficulty of testing, due to lack of explicit requirements for models. These challenges revealed a need for release checklists, i.e., formal lists of evaluation criteria that organizations should check either manually or (semi)automatically, to determine whether their product is ready for release to end users~\cite{breck2017ml,zinkevich2017rules,microsoftModelProduction}.

However, determining when generative AI-based software products are ready (safe) for release poses an even more complex challenge. LLMs do not only inherit the typical ML concerns around data dependency and model unpredictability, but they also face unique issues such as ensuring contextually accurate and unbiased language understanding, managing the vast and evolving scope of human language, and addressing the ethical implications of their responses in a wider range of real-world scenarios~\cite{achiam2023gpt}. Companies have already begun to share their processes and experiences through blogs and industry conferences, contributing to a growing body of ``grey literature'' on the subject.

This paper synthesizes a release checklist for generative AI-based products. Unlike traditional AI checklists, our checklist is compiled from $65$ grey literature sources across $44$ organizations, identifying key release challenges. Our results will enable future automation of release-readiness evaluation steps.

\section{BACKGROUND AND RELATED WORK}

Several studies have explored release-readiness for (traditional) ML software, drawing on industry experiences like Google's testing rubric~\cite{breck2017ml}, which emphasizes reliability and includes a 28-test scoring system for assessing ML system production readiness. Zinkevich~\cite{zinkevich2017rules} presents a Google-based guide on ML system development best practices, spanning from integration to feature engineering. Microsoft's production checklist~\cite{microsoftModelProduction} aids teams in evaluating ML model production readiness, focusing on performance, metrics, data quality, integration, and ethical considerations. Unlike these works, which target traditional ML, our study develops a release-readiness checklist specifically for generative AI-based software products.

\section{METHODOLOGY}

Our systematic grey literature review follows a similar approach for search construction as used by earlier work~\cite{raulamo2017choosing}, and aims at capturing the breadth of discussions surrounding the deployment of LLMs in production environments. We started from a base query, i.e., ``generative ai release checklist'', which covers the seminal resources in the field such as the ``LLMs in Production Conference''~\cite{mlopsLLMsProduction} and the influential blog posts of Matt Bornstein~\cite{bornstein2023} and Chip Huyen~\cite{huyenchipBuildingApplications}. We then split the query into three parts (1. ``generative ai'', 2. ``release'' and 3. ``checklist'') and included synonyms or related words to improve the coverage of grey literature, yielding the resulting search query:

\noindent\fbox{%
    \parbox{\linewidth}{%
        \texttt{("generative ai" | "large language models" | llms | "foundation models" | chatgpt | "conversational ai") AND (release | production | deployment | monitoring | observability | operations) AND (checklist | checks | guardrails | considerations | requirements | practices | patterns | challenges | methods | risks)}
    }%
}


Our search covered the period from June 1, 2018, the release date of GPT-1 by OpenAI, to September 1, 2023, the start of our survey.

\noindent\textbf{Inclusion criteria of our search:}

\begin{itemize}
    \item Literature must be authored by or associated with organizations focused on the field of generative AI.
    \item Content should specifically discuss LLMs in production settings.
    \item Accessibility to the full text without restrictions.
\end{itemize}

\noindent\textbf{The exclusion criteria:}

\begin{itemize}
    \item Product announcements, advertisements, and demos.
    \item News articles.
    \item Non-English literature.
    \item Duplicates.
    \item Literature that mentions LLM production without detailing the techniques used.
    \item Articles behind paywalls.
\end{itemize}

Our search yielded 522 million results, but relevance dropped after the top 100. We collected the top 100 results and augmented them with additional searches targeting the top 10 companies in generative AI and LLMs, as recognized by resources like State of AI~\cite{stateofStateReport} and Gartner~\cite{gartnerGenAI}. This approach netted 1,100 results, narrowed down to 35 through the above criteria, plus 30 more from snowball sampling.

The first and third authors developed a taxonomy for release-readiness challenges and solutions using a grey literature sample. Initially, they independently coded 20\% of the sample, then converged their taxonomies to form a consensus. This taxonomy was reapplied to re-code the initial sample, with their codings compared for consistency and conflicts resolved. Krippendorff's Kappa was calculated, resulting in a score of 0.917, indicating high inter-rater reliability ($>0.8$)~\cite{krippendorff2009testing}. This agreement allowed the first author to code the remaining 80\% of the sources. The findings were organized into a mindmap of release-readiness challenges and solutions, complemented by references. Subsequently, the first and second authors conducted workshops with their team to refine the mindmap into a checklist, incorporating team feedback.

Our replication package\footnote{https://github.com/SAILResearch/replication-24-harsh-generative-ai-release-readiness-checklist} includes the resulting mind map to visually organize the findings, the collected notes for each grey resource we reviewed and an online repository of references for further exploration. We refer to online references in this paper with the notation ``O<<number>>''.

\section{RESULTS: CHALLENGES}

\begin{figure*}[t]
    \centering
    \includegraphics[width=\textwidth]{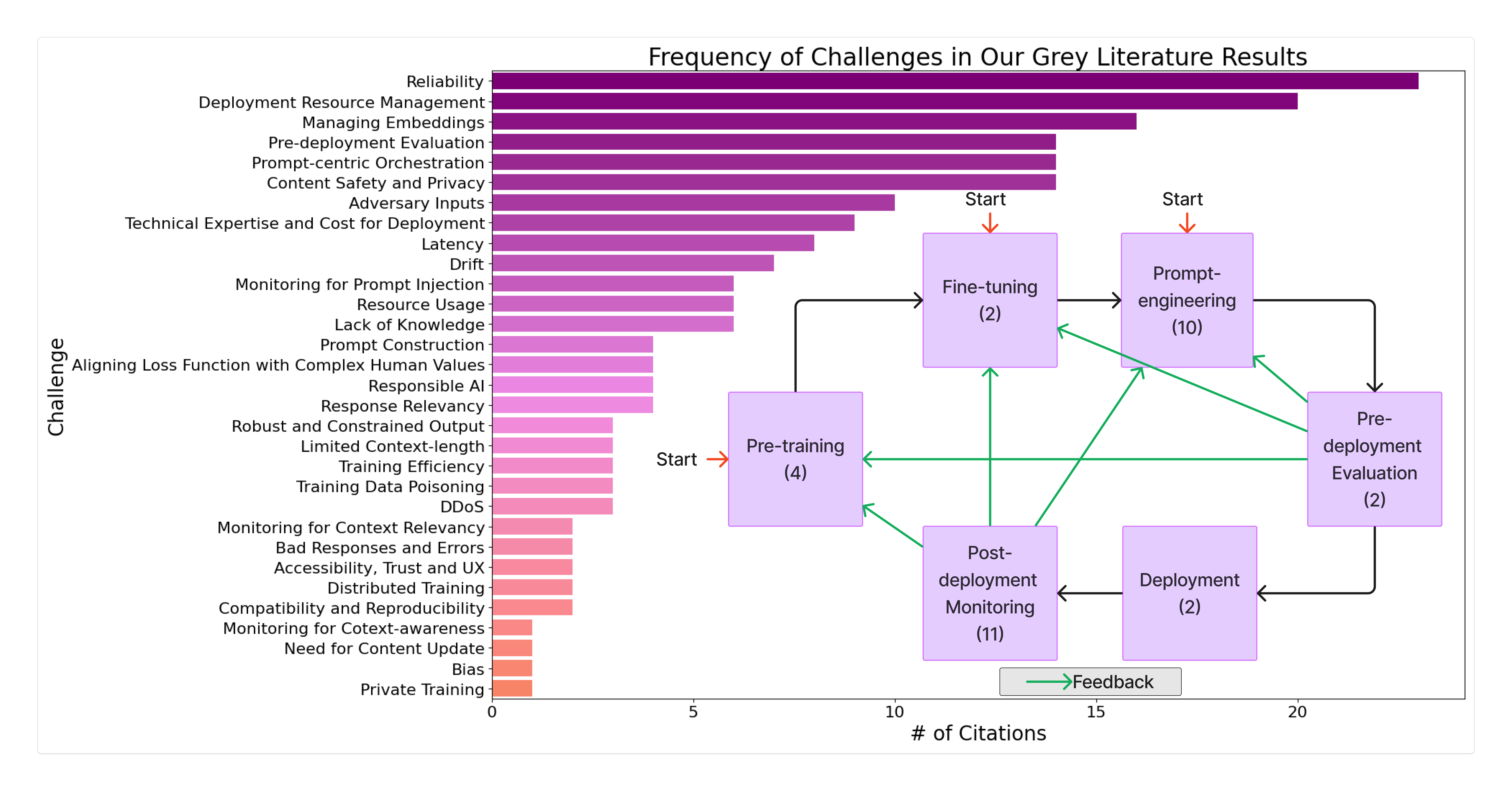}
    \caption{Frequency of challenges in our grey literature results (The numbers in the flowchart indicate the number of challenges associated with each stage of the LLM lifecycle)}
    \label{fig:challenges-histogram}
\end{figure*}

Figure \ref{fig:challenges-histogram} lists the 31 identified release challenges related to generative AI, and their prevalence in the grey sources. Note that the latter measures the number of ``mentions'' of these challenges in the grey sources, which is not necessarily equal to the actual frequency of challenges in practice (companies may use specific strategies or face challenges without necessarily discussing them in public media). The top five challenges discussed are:

\begin{enumerate}
    \item \textbf{Reliability:} Ensuring that LLMs consistently deliver accurate and coherent responses, avoiding hallucinations.
    \item \textbf{Deployment Resource Management:} Optimizing computational resources for LLMs in production to improve cost, latency, and performance.
    \item \textbf{Managing Embeddings:} Handling the creation, storage, and update of text documents' vector representations to improve semantic understanding and response relevance.
    \item \textbf{Pre-deployment Evaluation:} Conducting comprehensive testing to ensure LLMs' consistency, unbiasedness, and safety across demographics before release.
    \item \textbf{Prompt-centric Orchestration:} Effectively designing and managing prompts to utilize LLMs' full potential, addressing task decomposition, external tool integration, and sensitive information protection for privacy and security.
\end{enumerate}

\section{RESULTS: RELEASE-READINESS CHECKLIST}

This section presents a release-readiness checklist for LLM-based software, derived from our grey literature findings. It aligns with the software lifecycle stages depicted in Figure~\ref{fig:challenges-histogram}'s flowchart, highlighting key challenges per stage. The checklist offers checkpoints for assessing an LLM-based product's readiness and pinpoints improvement areas, including references and explanations of effective practices for each challenge. Practitioners should implement these practices or suitable adaptations for each relevant challenge.

\customsubsection{When pretraining a new LLM:}

\begin{enumerate}
    \item Check if your LLM needs to be trained using sensitive/private data provided by others. [O43]
          \newline \textbf{\textit{def.:}} Assess if LLM training requires third-party sensitive data, crucial for privacy compliance and ethical data use.
          \newline \textbf{\textit{sol.:}} Use Federated Learning (e.g., FedLLM, Nvidia Flare) to share models, not data, amongst partners by aggregating model parameters without compromising data privacy.
    \item Partition the training process across multiple accelerators (i.e., GPUs or Devices) to accelerate training. [O28][O42]
          \newline \textbf{\textit{sol.:}} Distributed platforms for LLM workloads, such as Ray, play a crucial role in enabling parallelization, particularly when incorporating advanced techniques like FL.
    \item Anonymize data. [O43]
          \begin{itemize}
              \item \textbf{\textit{def.:}} Alter or remove data to prevent association with individuals.
              \item \textbf{\textit{sol.:}} Use tools (Gretel.ai, Private AI, Tonic.ai) for data anonymization, complemented by manual checks to catch any nuances automation might miss.
          \end{itemize}
    \item Mitigate training data poisoning risks. [O26][O27][O29]
          \begin{itemize}
              \item \textbf{\textit{def.:}} Attackers may deliberately introduce misleading information in training datasets, leading LLMs to learn from biased and misleading data.
              \item \textbf{\textit{sol.:}} Check the source of external training data used by your organization, and keep records of its origins, similar to the methods used in the Software Bill of Materials (SBOM). Please see the SPDX specs\footnote{https://fossa.com/blog/spdx-3-0/,2023} for details about emerging AI and data SBOM scenarios.
          \end{itemize}
\end{enumerate}

\customsubsection{When fine-tuning an LLM:}

\begin{enumerate}
    \item Minimize unsafe behaviours or align loss function to complex human values. [O2][O25][O29][O58]
          \begin{itemize}
              \item \label{rlhf} \textbf{\textit{sol.:}} Reinforcement Learning with Human Feedback (RLHF) merges instruction fine-tuning with reinforcement learning, using human preferences (like pairwise comparisons) to train a reward model, aligning AI behavior with human values and safety. By doing so, RLHF aims to reduce the occurrence of unsafe behaviors, ensuring that AI actions are ethical, appropriate, and aligned with human expectations, thus enhancing the safety and reliability of AI applications.
          \end{itemize}
    \item Utilize efficient fine-tuning processes to improve training efficiency. [O2][O6][O9]
          \begin{itemize}
              \item \textbf{\textit{def.:}} With models growing in size, full fine-tuning is impractical on prosumer hardware, and storing and deploying separate fine-tuned models for each task is costly, as they are as large as the original model.
              \item \textbf{\textit{sol.:}} Parameter Efficient Fine-Tuning (PEFT) methods fine-tune only a few additional parameters, freezing most of the pre-trained LLM's parameters, significantly reducing computational and storage costs, and addressing catastrophic forgetting seen in full fine-tuning of LLMs.
          \end{itemize}
\end{enumerate}

\customsubsection{When doing prompt engineering:}

\begin{enumerate}
    \item Consider different prompt construction methods. [O4][O7][O10][O47]
          \newline \textbf{\textit{sol.:}}Explore Zero-shot, Few-shot, and Chain-of-Thought (CoT) prompting to significantly impact the effectiveness of LLMs. Crafting precise prompts, including clear instructions or questions and integrating contextual inputs or examples, leads to improved response quality and relevance.
    \item Utilize known practices to improve reliability and hallucinations. [O1][O2][O3][O4][O7][O9][O13][O14][O18][O19] [O22][O24][O26][O34][O36][O38][O40][O44][O46] [O47][O57][O60][O63]
          \begin{itemize}
              \item \textbf{\textit{def.:}} Achieving consistent outputs for identical inputs is crucial for user trust in (non-deterministic) LLMs.
              \item \textbf{\textit{sol.:}} Implement model-based evaluation or self-evaluation techniques, such as OpenAI Evals, which involve an LLM to critically self-assess its own outputs to ensure consistency, reliability, and ethical appropriateness.
              \item \textbf{\textit{sol.:}} Implement self-consistency and multiple prompting~\cite{wang2022self} by generating multiple responses for the same input and determining the final output through a majority vote or LLM selection. This can be done using popular LLM APIs like OpenAI's.
              \item \label{guardrails} \textbf{\textit{sol.:}} Guardrails ensure LLM outputs are coherent, accurate, factual, and free from harmful content, also safeguarding against adversarial inputs. Examples are Guardrails.ai, NeMo, Guidance, and rellm.
          \end{itemize}
    \item Bridge knowledge gaps by enriching prompts with relevant context. [O38][O47][O55][O56][O63]
          \begin{itemize}
              \item \textbf{\textit{def.:}} LLMs sometimes make incorrect assumptions to fill knowledge gaps, leading to hallucinations.
              \item \label{rag} \textbf{\textit{sol.:}} Utilize Retrieval-Augmented Generation (RAG) techniques to provide a richer context for prompts, thereby grounding model responses in factual content and mitigating hallucinations.
          \end{itemize}
    \item Utilize vector stores and embedding models to store, search and update vector representations of language data. [O2][O5][O7][O9][O10][O11][O14][O17][O33][O36] [O38][O44][O55][O60][O61][O63]
          \begin{itemize}
              \item \label{embeddings-and-vector-stores} \textbf{\textit{def.:}} Vector stores and embedding models are essential for managing and querying vector representations of language data, enabling efficient retrieval of relevant information (see \ref{rag}) and reducing hallucinations.
              \item \textbf{\textit{sol.:}} Vector stores like Weaviate, Pinecone, and pgvector, and embedding models by OpenAI, Huggingface and Cohere, are crucial for creating, storing, and updating vector representations of language data, facilitating more accurate and contextually relevant responses.
          \end{itemize}
    \item Employ efficient context-retrieval techniques to overcome LLM context length limit. [O1][O4][O38]
          \begin{itemize}
              \item \textbf{\textit{def.:}} Large models such as GPT-4 experience performance decline near their context window limit, affecting inference time and accuracy.
              \item \textbf{\textit{sol.:}} Implement chunking strategies to serve only the most relevant documents or segments for each prompt, enhancing the efficiency of information retrieval.
              \item \textbf{\textit{sol.:}} Apply contextual compression methods to condense and outline essential facts, thereby increasing the density of useful information within the LLM's context window.
          \end{itemize}
    \item If your application is used by untrusted users, you need an approach to deal with adversarial input. [O2][O4][O18][O19][O34][O26][O29][O40][O54][O57]
          \begin{itemize}
              \item \textbf{\textit{def.:}} LLMs require protection against prompts injected to manipulate model output.
              \item \textbf{\textit{sol.:}} Restrict a model with clear instructions within prompts as a fundamental defensive strategy.
              \item \textbf{\textit{sol.:}} Guardrails (see \ref{guardrails})
              \item \textbf{\textit{sol.:}} Adapt SQL injection defense strategies by parameterizing prompt components, thus separating instructions from inputs for enhanced security.
          \end{itemize}

    \item Ensure prompts' compatibility across changes to the underlying LLM (e.g., migrating from GPT3.5 to GPT4) and reproducibility across changes to the prompt itself. [O1][O19]
          \begin{itemize}
              \item \textbf{\textit{def.:}} In prompt engineering, unlike traditional software engineering where third-party library updates aim to ensure backward and/or forward compatibility after updates, there is no guarantee that prompts for an older model will work with a new version, often necessitating prompt rewrites. Therefore, if model changes are anticipated, it is vital to unit-test all prompts with evaluation examples.
              \item \textbf{\textit{sol.:}} Temperature ensures consistent output for the same input at the expense of less creativity.
              \item \textbf{\textit{sol.:}} Prompt versioning is essential for tracking the impact of minor changes on results, necessitating versioning and monitoring each prompt's performance.
              \item \textbf{\textit{sol.:}} Prompt templates provide a scalable way to create prompts, including instructions, few-shot examples, or action sequences for language models.
          \end{itemize}

    \item Incorporate practices to ensure content safety and privacy. [O1][O2][O4][O5][O9][O18][O24][O26][O27][O38] [O40][O57][O60]
          \begin{itemize}
              \item \textbf{\textit{def.:}} This check aims to ensure safety and privacy in third-party LLM-generated content, addressing issues that may remain post-RLHF (see \ref{rlhf}) fine-tuning. It involves reviewing and adjusting prompts to avoid generating harmful or sensitive information, ensuring compliance with ethical and privacy standards.
              \item \textbf{\textit{sol.:}} Models for detecting profanity, such as the PyPI module ``profanity-check''.
              \item \textbf{\textit{sol.:}} Use LLMs to evaluate responses for inappropriate content.
              \item \textbf{\textit{sol.:}} Use Personal Identifiable Information (PII) masking solutions with LangChain or LLamaIndex.
              \item \textbf{\textit{sol.:}} Utilize PII detection tools, such as Microsoft/presidio or Azure services.
          \end{itemize}

    \item Consider reusing prompts, decomposing a large task into smaller tasks, utilizing external tools or chaining prompts. [O7][O10][O11][O17][O36][O38][O40][O44][O55] [O61][O62]
          \begin{itemize}
              \item \textbf{\textit{def.:}} Prompt orchestration combines strategies to improve LLM effectiveness, such as reusing prompts, simplifying complex problems, incorporating external tools, and sequencing prompts for multi-step tasks.
              \item \textbf{\textit{sol.:}} Agent and chain orchestration frameworks like LangChain and LlamaIndex excel in simplifying prompt chaining, handling external API interactions, managing contextual data retrieval from vector stores, and preserving memory across multiple LLM calls.
              \item \textbf{\textit{sol.:}} Additionally, these frameworks leverage thought decomposition frameworks such as CoT, Tree of Thought, or ReAct to break down complex tasks into more manageable sub-tasks. This method not only simplifies the problem-solving process but also significantly enhances the effectiveness of the LLM in handling intricate tasks by guiding it through a structured approach to reasoning and response generation.
          \end{itemize}

    \item Ensure responses are highly constrained using types, templates and constraints. [O61][O62][O63]
          \newline \textbf{\textit{sol.:}} In systems combining LLMs with non-LLM components, it is crucial to maintain the response's clarity and usability. Constrained decoding, which enforces input constraints, lowers the risk of prompt injection attacks, boosting security (LMQL). Unlike guardrails, it limits LLM outputs at the token level to ensure each output step is controlled and valid.
\end{enumerate}

\customsubsection{Pre-deployment evaluation ensuring that a model's performance matches safety criteria for application deployment.}

\begin{enumerate}
    \item Evaluate the performance of your application pre-deployment. [O2][O5][O6][O9][O12][O15][O17][O19][O35][O36] [O38][O55][O56][O58]
          \begin{itemize}
              \item \textbf{\textit{def.:}} This involves measuring a model's performance against benchmarks and metrics to ensure it meets system and product standards. Such evaluations are essential for monitoring changes and identifying regressions without manual inspection of every model update.
              \item \textbf{\textit{sol.:}} Use A/B Testing to compare model responses and prompt designs in real-world scenarios, offering a direct insight into interaction quality.
              \item \textbf{\textit{sol.:}} For classification tasks with LLMs (like toxicity detection or document categorization) and extractive QA without dialogue, use standard metrics such as recall, precision, and PRAUC. For tasks lacking correct answers but having references, like machine translation or summarization, employ matching-based metrics (BLEU, ROUGE) or semantic similarity measures (BERTScore, MoverScore).
              \item \textbf{\textit{sol.:}} Conduct human evaluations to capture nuances in language that automated metrics may miss, providing a comprehensive assessment of model output quality.
              \item \textbf{\textit{sol.:}} Implement Penetration Tests and Red Teaming (i.e., ethical hackers identifying and exploiting vulnerabilities specific to LLM systems) to identify potential vulnerabilities in LLM systems, including biases or content generation issues, using resources like Anthropic's dataset for adversarial simulations\footnote{https://github.com/anthropics/hh-rlhf}.
              \item \textbf{\textit{sol.:}} Develop task-specific benchmarks with Eval Driven Development (EDD) for tasks like summarization or dialogue, focusing on metrics that reflect the unique aspects of each task, thus ensuring relevant and effective evaluations.
              \item \textbf{\textit{sol.:}} Utilize LLMs for self-evaluation using frameworks like G-Eval, incorporating Chain-of-Thought reasoning to enhance transparency and interpretability in assessments.
          \end{itemize}
    \item Consider practices to ensure accessibility and to build end-user trust in your application. [O2][O42]
          \newline \textbf{\textit{sol.:}} For comprehensive guidance on ensuring human-centric AI interaction, refer to the AI UI/UX guidelines provided by industry leaders such as Microsoft\footnote{https://www.microsoft.com/en-us/research/publication/guidelines-for-human-ai-interaction/}, Google\footnote{https://pair.withgoogle.com/guidebook/} and Apple\footnote{https://developer.apple.com/design/human-interface-guidelines/machine-learning}
\end{enumerate}

\customsubsection{During application deployment:}

\begin{enumerate}
    \item Check if you need a commercial or open-source LLM deployment. [O9][O19][O26][O27][O29][O33][O36][O37][O41]
          \begin{itemize}
              \item \textbf{\textit{def.:}} This decision parallels choosing between commercial and open-source software, considering ease of deployment, support, cost, versus flexibility and innovation potential.
              \item \textbf{\textit{sol.:}} Commercial LLMs simplify deployment and offer robust support, focusing on application development with a trade-off in higher costs and reduced flexibility.
              \item \textbf{\textit{sol.:}} Open-source LLMs require more technical and infrastructural investment but allow for greater customization and cost efficiency, supported by a vibrant community for development and troubleshooting.
          \end{itemize}

    \item Use optimization techniques to meet application latency, cost and resource requirements. [O2][O3][O4][O5][O9][O17][O19][O20][O24][O27] [O28][O32][O33][O36][O38][O41][O47][O60][O63] [O64]
          \begin{itemize}
              \item \textbf{\textit{def.:}} When choosing third-party LLMs, consider evolving costs and token size impacts. Opt between cloud scalability and on-premise security.
              \item \textbf{\textit{sol.:}} Semantic caching reduces latency and unnecessary computations by efficiently reusing previous responses, depending on the similarity algorithm used.
              \item \textbf{\textit{sol.:}} Methods like model compression, quantization, pruning, and distillation are used for model size reduction. Use them to decrease memory usage, improve computational efficiency, and lower latency.
              \item \textbf{\textit{sol.:}} Hardware configuration (GPUs, TPUs, CPUs) should be consistent with processing, memory, and storage needs for optimal performance.
              \item \textbf{\textit{sol.:}} Consider smaller, specialized LLMs for better cost-efficiency and resource management, often providing similar or superior performance for specific tasks.
              \item \textbf{\textit{sol.:}} Apply memory optimization and distributed inference strategies, such as data and tensor parallelism, to manage large models efficiently.
              \item \textbf{\textit{sol.:}} Leverage attention layer optimization methods like FlashAttention and PagedAttention to minimize memory requirements and enhance model responsiveness.
              \item \textbf{\textit{sol.:}} Optimize request scheduling to manage variable latency during inference, ensuring smooth user interactions with LLM-powered applications.
          \end{itemize}
\end{enumerate}

\customsubsection{Post-deployment monitoring to set up feedback loops and improve application performance.}

\begin{enumerate}
    \item Monitor for prompt injection attacks. [O7][O16][O18][O26][O38][O40]
          \begin{itemize}
              \item \textbf{\textit{def.:}} Detecting adversarial prompt inputs post-deployment to prevent manipulation of the LLM's outputs and avoid unintended interactions.
              \item \textbf{\textit{sol.:}} Utilize an adversarial prompt detector like "rebuff" to identify and filter adversarial prompts, leveraging LLMs' capabilities in specialized tasks like knowledge generation~\cite{liu2021generated} and self-verification~\cite{weng2022large}.
              \item \textbf{\textit{sol.:}} Analyze text similarity between known attacking prompts and current inputs to detect and mitigate potential threats.
          \end{itemize}

    \item Monitor the application for model resource usage. [O3][O9][O26][O32][O35][O63]
          \begin{itemize}
              \item \textbf{\textit{def.:}} Continuously assess model performance, resource consumption, and cost efficiency to fine-tune operations post-deployment.
              \item \textbf{\textit{sol.:}} Track LLM token count and utilization for cost-effective operation.
              \item \textbf{\textit{sol.:}} Regularly monitor system resources (CPU, GPU, memory usage) to maintain performance standards and adjust as necessary.
          \end{itemize}

    \item If your application is publicly exposed, consider solutions to prevent DDoS attacks. [O3][O8][O25]
          \begin{itemize}
              \item \textbf{\textit{sol.:}} Implement API rate limiting and use Captcha mechanisms to safeguard user experience and deter misuse by regulating access.
          \end{itemize}

    \item Monitor the application for model drift. [O1][O10][O16][O35][O38][O42] [O43]
          \begin{itemize}
              \item \textbf{\textit{def.:}} Addressing performance decline due to shifts in data distribution or user interaction patterns by monitoring input/output data.
              \item \textbf{\textit{sol.:}} Use historical performance data as a benchmark to identify and address drift in model behavior or data distribution.
              \item \textbf{\textit{sol.:}} Evaluate discrepancies between expected and actual prompts to adapt and refine LLM interactions based on real-world use.
              \item \textbf{\textit{sol.:}} Implement concept drift detection strategies and cluster analysis using embeddings (see \ref{embeddings-and-vector-stores}) to identify and correct drift issues post-deployment.
          \end{itemize}

    \item Check if your context-retrieval system can serve the "relevant" documents. [O14][O38]
          \begin{itemize}
              \item \textbf{\textit{def.:}} Ensuring the context-retrieval system matches user queries with relevant documents, especially in the face of unique or specific requests.
              \item \textbf{\textit{sol.:}} Measure query density to evaluate if the vector store accurately represents user queries, adjusting as needed to improve data relevance. Significant drift in query density indicates that the vector store lacks closely related data.
              \item \textbf{\textit{sol.:}} Utilize ranking metrics to assess and enhance the precision of the search and retrieval process, ensuring users receive the most pertinent information.
          \end{itemize}

    \item Evaluate the context served by the application to the LLM. [O58]
          \begin{itemize}
              \item \textbf{\textit{def.:}} Context evaluation in LLM applications is crucial for maintaining response credibility, referring to how the LLM uses the prompt, guiding information, and its inherent knowledge base to generate accurate and relevant responses.
              \item \textbf{\textit{sol.:}} Employ secondary LLMs for cross-evaluation of context relevance, quantifying the integrity of responses.
              \item \textbf{\textit{sol.:}} Critically evaluate how LLMs apply context in generating responses to ensure factual accuracy and appropriateness. The goal is to verify that the LLM correctly understands the topic, checks the accuracy of the information it references or provides, and incorporates this information in a way that is both correct and relevant.
          \end{itemize}

    \item Check LLM responses for relevance. [O16][O35][O56][O58]
          \begin{itemize}
              \item \textbf{\textit{def.:}} Regularly verify whether LLM outputs align with expected topics and maintain appropriate sentiment, w.r.t. the end user.
              \item \label{sentiment-score} \textbf{\textit{sol.:}} Track changes in LLM response sentiment to ensure consistency with expected topics, tone, and relevance for appropriate interactions.
              \item \textbf{\textit{sol.:}} Ensure responses stay relevant to predefined topics (e.g., politics).
              \item \textbf{\textit{sol.:}} Analyze the semantic similarity between queries and responses to ensure the LLM accurately addresses user intents.
          \end{itemize}

    \item Continuously evaluate the need for content updates for the context-retrieval system. [O38]
          \begin{itemize}
              \item \textbf{\textit{def.:}} Inability of the LLM to answer some queries may signal a need to update the vector store (see \ref{embeddings-and-vector-stores}).
              \item \textbf{\textit{sol.:}} Count an LLM's failures to answer prompts to create a refusal metric.
          \end{itemize}

    \item Ensure fairness. [O35][O40][O42][O43]
          \begin{itemize}
              \item \textbf{\textit{def.:}} LLMs can inherit and propagate biases from their training data. Organizations need to track and measure these biases in outputs using fairness metrics, which may vary by domain, including gender, race, or other unintentional biases.
              \item \textbf{\textit{sol.:}} Sentiment score (see \ref{sentiment-score})
              \item \textbf{\textit{sol.:}} Assess the model's performance across various demographic groups, e.g., Gender Bias can be measured by comparing the model's performance across different genders.
              \item \textbf{\textit{sol.:}} Combine toxicity classifiers with sentiment analysis (see \ref{sentiment-score}) to identify and mitigate harmful content in LLM outputs, ensuring they are unbiased, safe, and respectful.
          \end{itemize}

    \item Monitor the application for model latency. [O10][O11][O32][O35][O36] [O42][O57][O63]
          \begin{itemize}
              \item \textbf{\textit{def.:}} Keeping latency within acceptable limits is crucial for maintaining a positive user experience and operational efficiency in LLMOps.
              \item \textbf{\textit{sol.:}} Use observability tools to identify high-latency prompts via API latency metadata, allowing for targeted optimization.
          \end{itemize}

    \item Monitor the application for bad model responses. [O35][O40]
          \begin{itemize}
              \item \textbf{\textit{def.:}} LLMs can occasionally generate unwanted responses.
              \item \textbf{\textit{sol.:}} Value both implicit and explicit user feedback as key indicators for monitoring, especially negative or confused reactions.
          \end{itemize}
\end{enumerate}

\section{THREATS}

\subsection{External Validity}

Our checklist may not cover all nuances of LLM readiness due to the rapid advancements in the field since our survey. Despite this, the core challenges and solutions remain relevant. Consider the checklist a foundational tool for practitioners to build upon and adapt to the changing LLM technology landscape.

\subsection{Internal Validity}

Our checklist was created by a team of human coders, which may introduce bias. We attempted to mitigate this by using an established empirical methodology.

\section{CONCLUSION}

This paper investigates the release readiness of software products integrated with LLMs, synthesizing a comprehensive checklist to guide practitioners in evaluating their LLM products' readiness for release. As the generative AI landscape rapidly evolves, this checklist underscores the need for ongoing adaptation and community engagement to ensure the responsible and effective use of LLMs in software development.

\section{ACKNOWLEDGMENTS}

We are grateful to Mitacs and the National Bank of Canada for their support for our research.

\def\refname{REFERENCES}

\vspace*{-8pt}

\begin{IEEEbiography}{Harsh Patel}{\,} is currently pursuing a Master's degree at Queen's University, focusing on Software Engineering for Artificial Intelligence (SE4AI). His research primarily explores the release readiness of ML-based software products, with a special emphasis on post-deployment model recycling strategies. These strategies are designed to enable the efficient reuse of outdated models while maintaining their readiness for release. Harsh brings three years of software engineering experience to his research, having proficiency in technologies like Splunk, Python, JavaScript, Docker, and ReactJS. He is committed to advancing the integration of AI in software development processes. Contact him at patel.h@queensu.ca\vadjust{\vfill\pagebreak}
\end{IEEEbiography}

\begin{IEEEbiography}{Dominique Boucher, PhD}{\,} is currently Senior Director, Conversational AI Technologies at National Bank of Canada where he is responsible for the deployment of NBC dialogue systems. His main interests revolve around the use of conversational interfaces to help optimize business processes. He has been in the speech recognition and conversational AI industry for more than 25 years and holds a PhD from the University of Montreal.\vspace*{8pt}
\end{IEEEbiography}

\begin{IEEEbiography}{Emad Fallahzadeh, PhD}{\,} is a Postdoctoral Researcher at Queen's University in the Software Analysis and Intelligence Lab, specializing in mining software repositories. His research interests involve applying machine learning techniques and large language models to address various software engineering challenges. He is a member of the Association for Computing Machinery (ACM) and has contributed to numerous research endeavors in the field of software engineering, including publications in the International Conference on Software Engineering (ICSE) and the Foundations of Software Engineering (FSE) conference proceedings. Contact him at emad.fallahzadeh@queensu.ca\vspace*{8pt}
\end{IEEEbiography}

\begin{IEEEbiography}{Ahmed E. Hassan, PhD} {\,} is the NSERC/RIM Industrial Research Chair in Software Engineering for Ultra Large Scale systems at Queen's University, Canada. He spearheaded the organization and creation of the Mining Software Repositories (MSR) Conference and its research community. He co-edited special issues of the IEEE Transactions on Software Engineering and the Journal of Empirical Software Engineering on the MSR topic. Early tools and techniques developed by his team are already integrated into products used by millions of users worldwide. His industrial experience includes helping architect the Blackberry wireless platform at RIM, and working for IBM Research at the Almaden Research Lab and the Computer Research Lab at Nortel Networks. He is the named inventor of patents in several jurisdictions around the world, including the United States, Europe, India, Canada, and Japan. He is a member of the IEEE.\vspace*{8pt}
\end{IEEEbiography}

\begin{IEEEbiography}{Bram Adams, PhD}{\,} is a full professor at Queen's University. His research interests include software release engineering (pre- and post-AI) and mining software repositories. His work has received the 2021 Mining Software Repositories Foundational Contribution Award. In addition to co-organizing the RELENG International Workshop on Release Engineering from 2013 to 2015 (and the 1st/2nd IEEE Software Special Issue on Release Engineering), he co-organized the first editions of the SEMLA event on Software Engineering for Machine Learning Applications. He has been PC co-chair of SCAM 2013, SANER 2015, ICSME 2016 and MSR 2019, and ICSE 2023 software analytics area co-chair. He is a Senior IEEE Member. Contact him at bram.adams@queensu.ca\vspace*{8pt}
\end{IEEEbiography}


\begin{thebibliography}{1}

    \bibitem{hou2023large}Hou, X., Zhao, Y., Liu, Y., Yang, Z., Wang, K., Li, L., Luo, X., Lo, D., Grundy, J. \& Wang, H. Large language models for software engineering: A systematic literature review. {\em ArXiv Preprint ArXiv:2308.10620}. (2023)

    \bibitem{breck2017ml}Breck, E., Cai, S., Nielsen, E., Salib, M. \& Sculley, D. The ML test score: A rubric for ML production readiness and technical debt reduction. {\em 2017 IEEE International Conference On Big Data (Big Data)}. pp. 1123-1132 (2017)

    \bibitem{zinkevich2017rules}Zinkevich, M. Rules of machine learning: Best practices for ML engineering. {\em URL: Https://developers. Google. Com/machine-learning/guides/rules-of-ml}. (2017)

    \bibitem{microsoftModelProduction}ISE, M. ML model production checklist - Engineering Fundamentals Playbook — microsoft.github.io. (https://microsoft.github.io/code-with-engineering-playbook/machine-learning/ml-model-checklist/,0), [Accessed 21-02-2024]

    \bibitem{mlopsLLMsProduction}Community, M. LLMs in Production Conference - Event | MLOps Community — home.mlops.community. (https://home.mlops.community/public/events/llms-in-production-conference-2023-04-13,0), [Accessed 21-02-2024]

    \bibitem{huyenchipBuildingApplications}Huyen, C. Building LLM applications for production — huyenchip.com. (https://huyenchip.com/2023/04/11/llm-engineering.html,0), [Accessed 21-02-2024]

    \bibitem{wang2022self}Wang, X., Wei, J., Schuurmans, D., Le, Q., Chi, E., Narang, S., Chowdhery, A. \& Zhou, D. Self-consistency improves chain of thought reasoning in language models. {\em ArXiv Preprint ArXiv:2203.11171}. (2022)

    \bibitem{liu2021generated}Liu, J., Liu, A., Lu, X., Welleck, S., West, P., Bras, R., Choi, Y. \& Hajishirzi, H. Generated knowledge prompting for commonsense reasoning. {\em ArXiv Preprint ArXiv:2110.08387}. (2021)

    \bibitem{weng2022large}Weng, Y., Zhu, M., He, S., Liu, K. \& Zhao, J. Large language models are reasoners with self-verification. {\em ArXiv Preprint ArXiv:2212.09561}. (2022)


    \bibitem{stateofStateReport}Benaich, N. \& Capital, A. State of AI Report 2023 — stateof.ai. (https://www.stateof.ai/,0), [Accessed 28-02-2024]

    \bibitem{gartnerGenAI}Gartner Generative AI: What Is It, Tools, Models, Applications and Use Cases. (https://www.gartner.com/en/topics/generative-ai,0), [Accessed 28-02-2024]

    \bibitem{achiam2023gpt}Achiam, J., Adler, S., Agarwal, S., Ahmad, L., Akkaya, I., Aleman, F., Almeida, D., Altenschmidt, J., Altman, S., Anadkat, S. \& Others Gpt-4 technical report. {\em ArXiv Preprint ArXiv:2303.08774}. (2023)

    \bibitem{raulamo2017choosing}Raulamo-Jurvanen, P., Mäntylä, M. \& Garousi, V. Choosing the right test automation tool: a grey literature review of practitioner sources. {\em Proceedings Of The 21st International Conference On Evaluation And Assessment In Software Engineering}. pp. 21-30 (2017)

    \bibitem{bornstein2023}Bornstein, M. \& Radovanovic, R. Emerging Architectures for LLM Applications. (https://a16z.com/emerging-architectures-for-llm-applications/,2023), [Accessed 01-03-2024]


    \bibitem{krippendorff2009testing}Krippendorff, K. Testing the reliability of content analysis data. {\em The Content Analysis Reader}. pp. 350-357 (2009)

\end{thebibliography}
\end{document}